\documentclass[11pt]{article}

\usepackage[a4paper,margin=1in]{geometry}
\usepackage{amsfonts,amssymb,amsmath,amsthm}
\usepackage{graphicx}
\usepackage{psfrag}
\usepackage{calc}
\usepackage{epic}
\usepackage[english]{babel}
\usepackage{hyperref}
\usepackage{float}
\usepackage{latexsym}
\usepackage{color}
\usepackage{subfigure,dsfont}

\newtheorem{theorem}{Theorem}
\newtheorem{lemma}[theorem]{Lemma}

\newcommand{\be}{ \begin{equation}}
\newcommand{\ee}{\end{equation}}
\newcommand{\ben}{ \begin{equation*}}
\newcommand{\een}{\end{equation*}}

\newcommand{\J}{\mathcal{J}}

\def\E{{\mathbb E}}

\def\I{{\mathbf I}}
\def\W{{\mathbf W}}
\def\Q{{\mathbf Q}}
\def\l{\lambda}
\def\m{\mu}

\DeclareMathOperator*{\argmax}{arg\,max}

\usepackage{hyperref}
\newcommand{\footremember}[2]{%
    \footnote{#2}
    \newcounter{#1}
    \setcounter{#1}{\value{footnote}}%
}

\title{Static vs accumulating priorities \\ in healthcare queues under heavy loads}

\author{%
Binyamin Oz\footremember{HU}{Hebrew University of Jerusalem}%
 \and Seva Shneer\footremember{HW}{Heriot-Watt University}%
  \and Ilze Ziedins\footremember{UA}{The University of Auckland}%
  }
\date{}

\begin{document}

\maketitle

\abstract{Amid unprecedented times caused by COVID-19, healthcare systems all over the world are strained to the limits of, or even beyond, capacity. A similar event is experienced by some healthcare systems regularly, due to for instance seasonal spikes in the number of patients. We model this as a queueing system in heavy traffic (where the arrival rate is approaching the service rate from below) or in overload (where the arrival rate exceeds the service rate). In both cases we assume that customers (patients) may have different priorities and we consider two popular service disciplines: static priorities and accumulating priorities. It has been shown that the latter allows for patients of all classes to be seen in a timely manner as long as the system is stable. We demonstrate however that if accumulating priorities are used in the heavy traffic or overload regime, then all patients, including those with the highest priority, will experience very long waiting times. If on the other hand static priorities are applied, then one can ensure that the highest- priority patients will be seen in a timely manner even in overloaded systems.}

\section{Introduction}

We are currently seeing the effect COVID-19 has on healthcare services in vast majority of countries in the world. Healthcare services are also under increasing pressure as demographics change and populations age.  Public health services are struggling, and sometimes failing, to maintain services under this increasing load.  Given this context of high demand for tightly constrained resources it is instructive to reassess the rationales for prioritization regimes currently in use, and contrast with other possibilities.  Specifically, in this paper we characterize the performance of the very commonly used static priority regime, and contrast it with the recently proposed accumulating priority regime, under critical loadings.

Healthcare systems have traditionally used static priority queues in a range of settings from triage in an emergency departments (EDs), to organizing access to elective surgeries, such as hip and knee replacements \cite{dingetal2019EDtriage},\cite{arnettetal2003hipknee}. In a static priority regime, patients are assigned to a priority class, and must wait to be treated until all patients in higher priority classes have been treated.  Patients may sometimes be moved to a higher priority class if their condition deteriorates, but in practice there is often no automatic mechanism for making such transitions, and any adjustments may rely on patients proactively approaching their healthcare provider.  Accumulating priority queues (APQs) have recently been proposed to overcome some of the inherent drawbacks of static priority queues in healthcare \cite{stanford2014waiting}.  In accumulating priority queues, patients accumulate priority with time spent in the queue, at a rate that depends on their priority class, with higher priority patients accumulating priority faster than lower priority patients.  Priority can accumulate linearly, or in a nonlinear fashion.  Observational studies of behaviour in emergency departments have revealed that in practice physicians may operate a regime that is similar to an APQ, with the likelihood of being seen increasing more rapidly as waiting times approach threshold targets, see e.g.\ \cite{dingetal2019EDtriage}.

The accumulating priority regime was first proposed by Kleinrock \cite{kleinrock1964delay}, who obtained expressions for the expected waiting times for all classes.  A large-deviations principle has been established in \cite{stolyar2001largest}. More recently, Stanford et al.\ \cite{stanford2014waiting} derived expressions for the Laplace Stieltjes Transform of the waiting time distribution, which can then be inverted numerically. A later paper \cite{lietal2017nonlinear}, showed that a wide range of possible accumulation functions (including, for instance, exponential and log) have an equivalent linear regime, in the sense that the order in which patients are seen is the same in both the nonlinear and linear formulation. 

This paper considers the performance of a single server queue with total arrival rate $\rho$ and service rate 1, where 
$\rho$, the load on the server, either satisfies $\rho \uparrow 1$ or $\rho > 1$. The heavy traffic regime has been intensively studied, although not for the accumulating priority queue.   When $\rho > 1$ queues are overloaded and hence unstable, and no equilibrium exists.  Unstable queues, if unchecked, grow without bound, which is unrealistic for almost every application, and of course, an infinite queue never exists in practice.  However, there are many applications where arrival rates are greater than service rates for shorter or longer periods.  Traffic networks are an immediate example.  In many cities rush hour queues build up, and then decay, not because the system can cope with the increased volume of traffic but simply because the flow into the network has reduced as rush hour passes. 

At the time of writing, one cannot overstate the importance of looking at healthcare systems in the situations when demand suddenly exceeds capacity. We see healthcare systems of a very large number of countries being overwhelmed with an influx of patients. 

Healthcare can however suffer from a similar excess of demand over capacity in other situations, particularly in winter, and for those healthcare systems which experience loss of capacity over longer periods, addressing the issue of how best to organize patient prioritization is vital \cite{NHSworkforce}.
Emergency departments may also see increased demand due to patients who could have been treated by their GP, but were deterred by cost (e.g.\ in New Zealand, where hospital visits in the public health system are free, but GP visits are not), or long waiting times for GP appointments.  No patient arriving in a period when $\rho > 1$ actually sees an infinite queue -- rather they see a large, possibly very large, queue ahead of them.   Their waiting time for treatment, given they have joined the queue, is also not infinite, but as a patient's condition deteriorates, it may nevertheless be far too long.  In practice also, if $\rho$ is close to $1$, then over relatively short periods of time it may be difficult to determine whether $\rho <1$ or $\rho >1$. Thus there is a strong practical need to address the question of how best to organize patient prioritization in this transient regime.

Indeed, we will see below that the two cases: a) $\rho << 1$ and b) $\rho \uparrow 1$ or $\rho > 1$, require fundamentally different approaches to patient prioritization. If $\rho << 1$, then the accumulating priority regime ensures that all patients are seen in a timely manner,  while ensuring health targets are met (if those targets are feasible). On the other hand, if $\rho \uparrow 1$ or $\rho > 1$, then it is impossible to limit waiting times for all patients, and the static priority regime provides a mechanism for ensuring that healthcare is still available to the most acute, and vulnerable, patients (see Fig. \ref{fig:static} below), whereas under APQ the expected waiting times for all classes increase (Fig. \ref{fig:acc} below).  For this scenario we propose below a mixture of the two prioritization schemes.

Section 2 gives a detailed description of the models we consider.  Section 3 considers the case where $\rho = 1 - \epsilon$, as $\epsilon \downarrow 0$, while Section 4 considers the case where $\rho > 1$.  In Section 5 we discuss issue related to customers making strategic decisions. 
We conclude with a short discussion of related models and potential future research directions in Section 6.

\section{Model}

We consider a service system with a single server and $N+1$ classes of customers. Customers (patients) of class $i$ arrive according to a Poisson process with rate $\lambda_i$ and their priority class is associated with a positive real number $b_i$ for $1 \leq i \leq N+1$.  The higher the number $b_i$, the higher the priority class, and without loss of generality we assume $b_1 > b_2 > \ldots > b_N > b_{N+1}$. Thus arrivals of class 1 are in the highest priority class and arrivals of class $N+1$ are in the lowest priority class.

If a customer finds the server idle upon its arrival, the server immediately starts serving this customer. Priority is non-preemptive, that is, if the server is busy when a new customer arrives, the customer joins the waiting room regardless of her priority class. The room is assumed to be of infinite size.
Whenever the server finishes serving a customer and the waiting room is non-empty, the server starts to serve the customer with the highest current priority among those currently in the waiting room.

We consider two different priority policies:
\begin{itemize}
\item{}
{\bf Static priorities (SP)}: in this case a customer of class $i$ has priority level $b_i$ which does not change;
\item{}
{\bf Accumulating priorities (AP)}: in this case a customer of class $i$ that spent $s$, $s\ge 0$, units of time waiting in the waiting room has priority level $b_i s$.
\end{itemize}

We assume, without loss of generality, that service times for customers are independent and exponentially distributed with mean $1$. Therefore
\begin{equation} \label{eq:def_rho}
\rho = \lambda_1 + \ldots + \lambda_{N+1}
\end{equation}
is the load on the system -- the average amount of new work arriving per unit of time -- and $\rho < 1$ is necessary for stability.


We will consider the system in high loads in two scenarios. In the first scenario we assume that $\rho = 1-\varepsilon$ and $\varepsilon \downarrow 0$. The system is stable for any $\varepsilon > 0$, but, as $\varepsilon$ decreases to $0$, the average number of customers waiting in the queue in the stationary regime (and, by Little's formula, the average waiting time of a typical customer arriving in the system) tends to infinity. In the second scenario $\rho > 1$ and the system is thus unstable.

Our goal is to study the behavior of the expected waiting time of a customer from each of the classes, in the two scenarios described above, and under the two different priority policies.

\section{Loads approaching capacity}

In this section we consider a sequence of systems indexed by $\varepsilon > 0$ such that the loads in these systems increase to $1$. More precisely, we assume that $\lambda_1(\varepsilon), \ldots, \lambda_{N+1}(\varepsilon)$ are non-decreasing functions of $\varepsilon > 0$ such that $\lambda_i(\varepsilon) \uparrow \lambda_i$ as $\varepsilon \downarrow 0$ for all $i=1,\ldots, N+1$, 
$$
\sum_{i=1}^{N+1} \lambda_i(\varepsilon) = 1 - \varepsilon
$$
for all $\varepsilon > 0$, and
$$
\sum_{i=1}^{N+1} \lambda_i = 1.
$$
An interesting special case of the setting above is the scenario where $\lambda_1,\ldots, \lambda_N$ are fixed and $\lambda_{N+1}(\varepsilon) = 1-(\lambda_1+\ldots+\lambda_N)-\varepsilon$, but we do not restrict ourselves to this case.

Regardless of the service discipline chosen, the systems are stable for any $\varepsilon > 0$. Denote by $(Q_1^\varepsilon, \ldots, Q_{N+1}^{\varepsilon})$ the vector of steady-state queue lengths and by $(W_1^\varepsilon, \ldots, W_{N+1}^{\varepsilon})$ the vector of steady-state waiting times (inclusive of the service time), for a particular value of $\varepsilon$. We also let $\Q_i^\varepsilon = \E\left(Q_i^\varepsilon\right)$ and  $\W_i^\varepsilon = \E\left(W_i^\varepsilon\right)$ be the expected queue length and waiting times respectively, for all $i$. Regardless of the service discipline,
\begin{equation} \label{eq:ht_total_queue}
\Q_1^\varepsilon + \ldots + \Q_{N+1}^\varepsilon = \frac{1-\varepsilon}{\varepsilon} \sim \frac{1}{\varepsilon}
\end{equation}
as $\varepsilon \to 0$. The total queue lengths thus increase to infinity, and we are interested in how queue lengths, and waiting times, of the individual classes behave.

\subsection{Static priorities}

Consider first the {\bf SP} priorities.  Let $\sigma_i(\varepsilon) = \sum_{k=1}^i \lambda_k(\varepsilon), 1 \leq i \leq N+1$, with $\sigma_0(\varepsilon) = 0.$ Let also $\sigma_i = \lim_{\varepsilon \downarrow 0} \sigma_i(\varepsilon)$.  From Cobham \cite{cobham1954staticpriority}, we obtain that the expected waiting times for the priority classes are given by
\begin{equation}
\W^\varepsilon_i = \frac{1 - \varepsilon}{(1 - \sigma_{i-1}(\varepsilon))(1 - \sigma_i(\varepsilon))} + 1, ~~~~~~~~~~1 \leq i \leq N+1.
\end{equation}
and we can write 
\begin{eqnarray*}
\W^\varepsilon_{N+1} & = & \frac{\frac{1}{\varepsilon} - 1}{\varepsilon + \lambda_{N+1}(\varepsilon)} + 1.
\end{eqnarray*}
Thus, we see that as $\varepsilon \rightarrow 0$,
\begin{eqnarray*}
\W^\varepsilon_i & \rightarrow  & \frac{1}{(1 - \sigma_{i-1})(1 - \sigma_i)} + 1 < \infty , ~~~~~~~~~~1 \leq i \leq N. \\[1mm]
\W^\varepsilon_{N+1}  & \rightarrow & \infty
\end{eqnarray*}
and 
\begin{eqnarray*}
\Q_i^\varepsilon & <& \infty, ~~~~~~~~~~~1 \leq i \leq N \\
\Q_{N+1}^\varepsilon &\sim& 1/\varepsilon.
\end{eqnarray*}
Thus, as $\varepsilon \rightarrow 0$, in the {\bf (SP)} case,  the expected queue lengths and waiting times for classes 1 to $N$ are bounded from above, while for class $N+1$ they grow without bound.  The durations of busy periods also
increase without bound.


\subsection{Accumulating priorities}

For the {\bf AP} case we can conclude from the Kleinrock formula (\cite{kleinrock1964delay}, see also \cite{stanford2014waiting}) that waiting times for all customers grow without bound as $\varepsilon \rightarrow 0$, 
so that if the $b_i$ are held
constant, this regime does not offer the same protection for the higher priority classes as {\bf SP} does.
Indeed we can prove the following exact statement.

\begin{lemma} \label{lem:ap_ht}
Consider an accumulating priority queue with $N+1$ classes, and accumulation rates $b_1 > b_2 > \ldots > b_{N+1}$. Then
$$
\lim_{\varepsilon \downarrow 0} \varepsilon \W_i^\varepsilon = \frac{1/b_i}{\sum\limits_{k=1}^{N+1} \lambda_k/b_k}, ~~~~~~~~1 \leq i \leq N+1.
$$
\end{lemma}

We present a proof of Lemma \ref{lem:ap_ht} below but first comment on its implications. The result may be interpreted as follows: a customer from class $i$ entering service after waiting for time $W_i$ has, at that time, priority $b_i W_i$. Lemma \ref{lem:ap_ht} essentially states that the {\bf (AP)} discipline makes all these priorities just before service equal on average, across classes, in heavy traffic.  This is similar to the behaviour in heavy traffic of the MaxWeight protocol (see \cite{stolyar2004maxweight}) which equates scaled queue lengths (we discuss this connection further below).

Lemma \ref{lem:ap_ht} also implies that $\W_i^\varepsilon \to \infty$ as $\varepsilon \downarrow 0$ for all $i$, and hence
\begin{equation} \label{eq:ht_result_queues}
\lim_{\varepsilon \downarrow 0} \varepsilon \Q_i^\varepsilon = \lim_{\varepsilon \downarrow 0} \varepsilon \lambda_i(\varepsilon) \W_i^\varepsilon = \frac{\lambda_i/b_i}{\sum\limits_{k=1}^{N+1} \lambda_k/b_k},
\end{equation}
which also means that
\begin{equation} \label{eq:ht_result_queues}
\lim_{\varepsilon \downarrow 0} \frac{\Q_i^\varepsilon}{\sum_{k=1}^{N+1} \Q_k^\varepsilon} = \frac{\lambda_i/b_i}{\sum\limits_{k=1}^{N+1} \lambda_k/b_k}.
\end{equation}

{\bf Proof of Lemma \ref{lem:ap_ht}.} In order to simplify notation, in this proof we will drop the dependence of $\lambda_i$ and $\W_i$ on $\varepsilon$. The formulas from \cite{kleinrock1964delay} adapted to our setting are as follows:
$$
\W_i = \frac{1/\varepsilon - 1 - \sum_{k=i+1}^{N+1} \lambda_k \left(1-\frac{b_k}{b_i}\right) \W_k}{1-\sum_{k=1}^i \lambda_k \left(1-\frac{b_i}{b_k}\right)}.
$$
We use a proof by induction on $i=N+1,\ldots, 1$. First write
\begin{align*}
\W_{N+1} & = \frac{1/\varepsilon - 1}{1-\sum_{k=1}^{N+1} \lambda_k \left(1-\frac{b_{N+1}}{b_k}\right)} = \frac{1/\varepsilon - 1}{1-\sum_{k=1}^{N+1} \lambda_k + b_{N+1} \sum_{k=1}^{N+1} \lambda_k/b_k} \\ & =  \frac{1/\varepsilon - 1}{\varepsilon + b_{N+1} \sum_{i=1}^{N+1} \lambda_k/b_k}, 
\end{align*}
which implies the statement of the lemma for $i=N+1$. Assume now the statement is valid for all $i \ge j+1$ and let us prove if for $i=j$:
\begin{align*}
\lim_{\varepsilon \downarrow 0} \varepsilon \W_j & = \frac{1}{1-\sum_{k=1}^j \lambda_k \left(1-\frac{b_j}{b_k}\right)} \left(1- \sum_{k=j+1}^{N+1} \lambda_k \left(1-\frac{b_k}{b_j}\right) \lim_{\varepsilon \downarrow 0} \varepsilon \W_k\right) \\
& = \frac{1}{1-\sum_{k=1}^j \lambda_k + b_j \sum_{k=1}^j \lambda_k/b_k} \left(1-\frac{\sum_{k=j+1}^{N+1} \lambda_k \left(1-\frac{b_k}{b_j}\right) 1/b_k}{\sum_{k=1}^{N+1} \lambda_k/b_k}\right) \\
& = \frac{1}{\sum_{k=1}^{N+1} \lambda_k/b_k} \frac{1}{1-\sum_{k=1}^j \lambda_k + b_j \sum_{k=1}^j \lambda_k/b_k} \left(\sum_{k=1}^{N+1} \lambda_k/b_k -  \sum_{k=j+1}^{N+1} \lambda_k/b_k + \frac{1}{b_j} \sum_{k=j+1}^{N+1} \lambda_k\right) \\
& = \frac{1}{\sum_{k=1}^{N+1} \lambda_k/b_k} \frac{\sum_{k=1}^{j} \lambda_k/b_k + \frac{1}{b_j} \sum_{k=j+1}^{N+1} \lambda_k}{\sum_{k=j+1}^{N+1} \lambda_k + b_j \sum_{k=1}^j \lambda_k/b_k} = \frac{1/b_j}{\sum_{k=1}^{N+1}\lambda_k/b_k}.
\end{align*}
\qed

\subsection{Solution}

These results suggest that in the {\bf AP} regime, the accumulation rates need to be adjusted if the system is subjected 
to increasingly heavy loads.  A natural solution we propose is to take $b_i(\varepsilon) = c_i/\varepsilon$ for fixed $c_i$,  $\{1,\ldots,N\}$, with $b_{N+1} = c_{N+1}$.  This effectively applies the {\bf SP} regime to the lowest priority class, while maintaining the positive benefits of the {\bf AP} regime for the remaining classes.  More generally, we could consider regimes where $b_i(\varepsilon) = c_i/\varepsilon$ for fixed $c_i$,  $\{1,\ldots,M\}$ for any $M \leq N$.  If $M < N$, then the split between classes following {\bf AP} and those following {\bf SP} will still provide benefits to the higher priority classes.  Whatever the split, only the lowest priority class will experience delays that grow without bound.

\subsection{Numerical illustrations}

We illustrate the conclusions above with typical sample paths of a system with $N=2$ (there are therefore $3$ classes of customers) under different priority regimes. Assume that arrival intensities are given by $\lambda_1 = \lambda_2 = 1/3$ and $\lambda_3 = 1/3-\varepsilon$, and priorities $b_1=3$, $b_2=2$ and $b_3=1$. 

\begin{figure}
\centering
\includegraphics[width=\linewidth, height=0.35\linewidth]{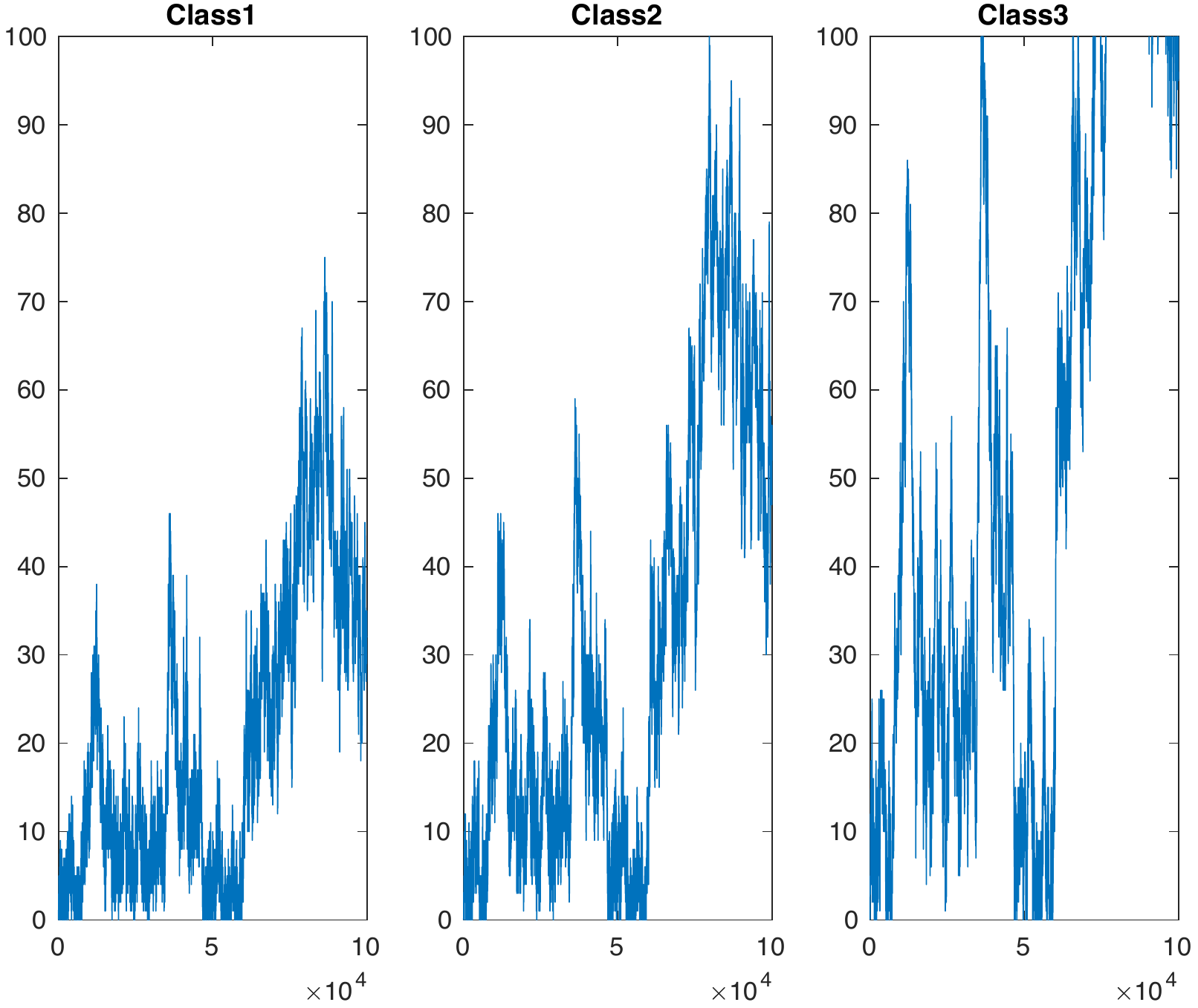}
\caption{Numbers of customers of different classes against time in the system with accumulating priorities, with $\varepsilon=0.001$}
\label{fig:acc}
\end{figure}

\begin{figure}
\centering
\includegraphics[width=\linewidth, height=0.35\linewidth]{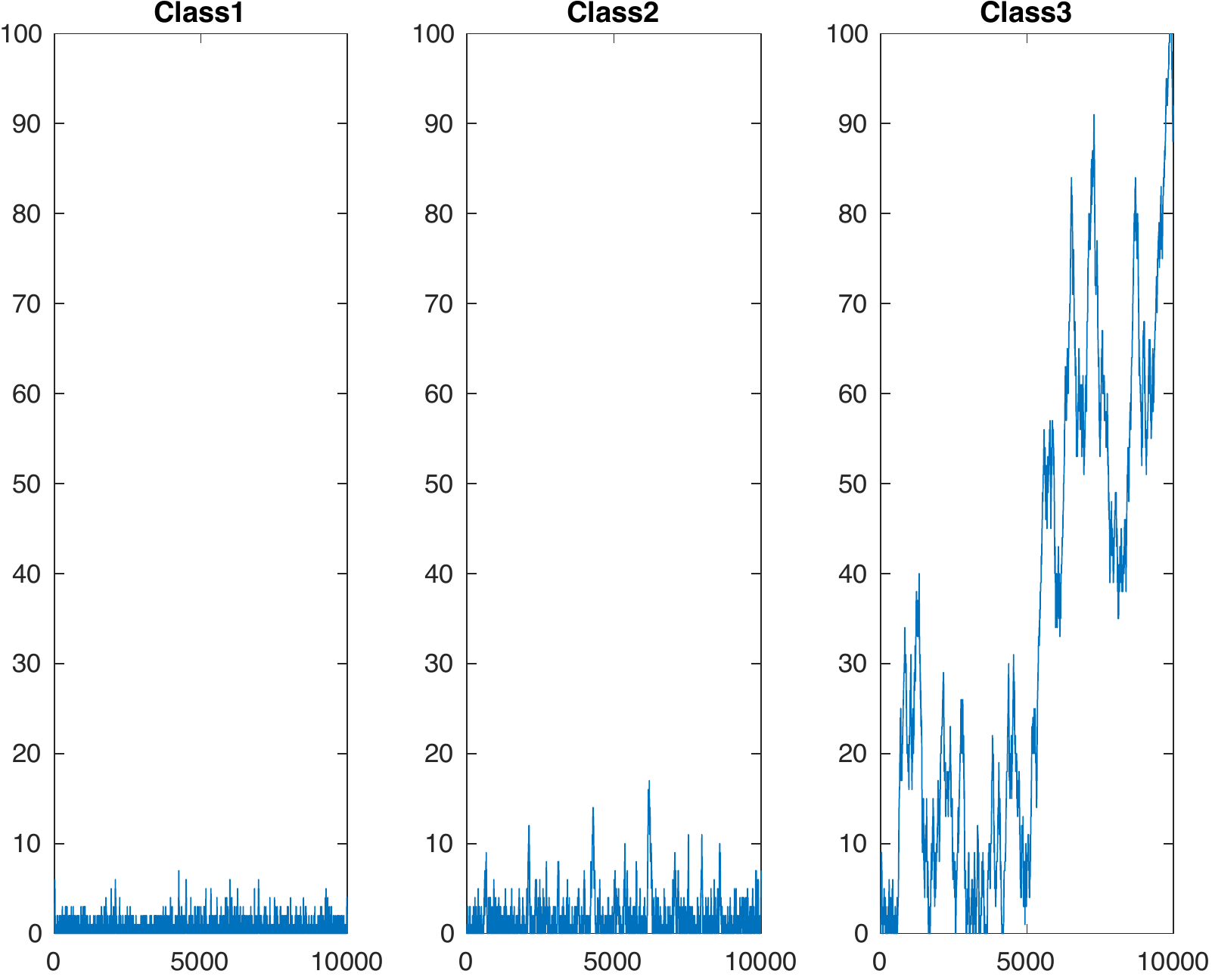}
\caption{Numbers of customers of different classes against time in the system with static priorities, with $\varepsilon=0.001$}
\label{fig:static}
\end{figure}

\begin{figure}
\centering
\includegraphics[width=\linewidth, height=0.35\linewidth]{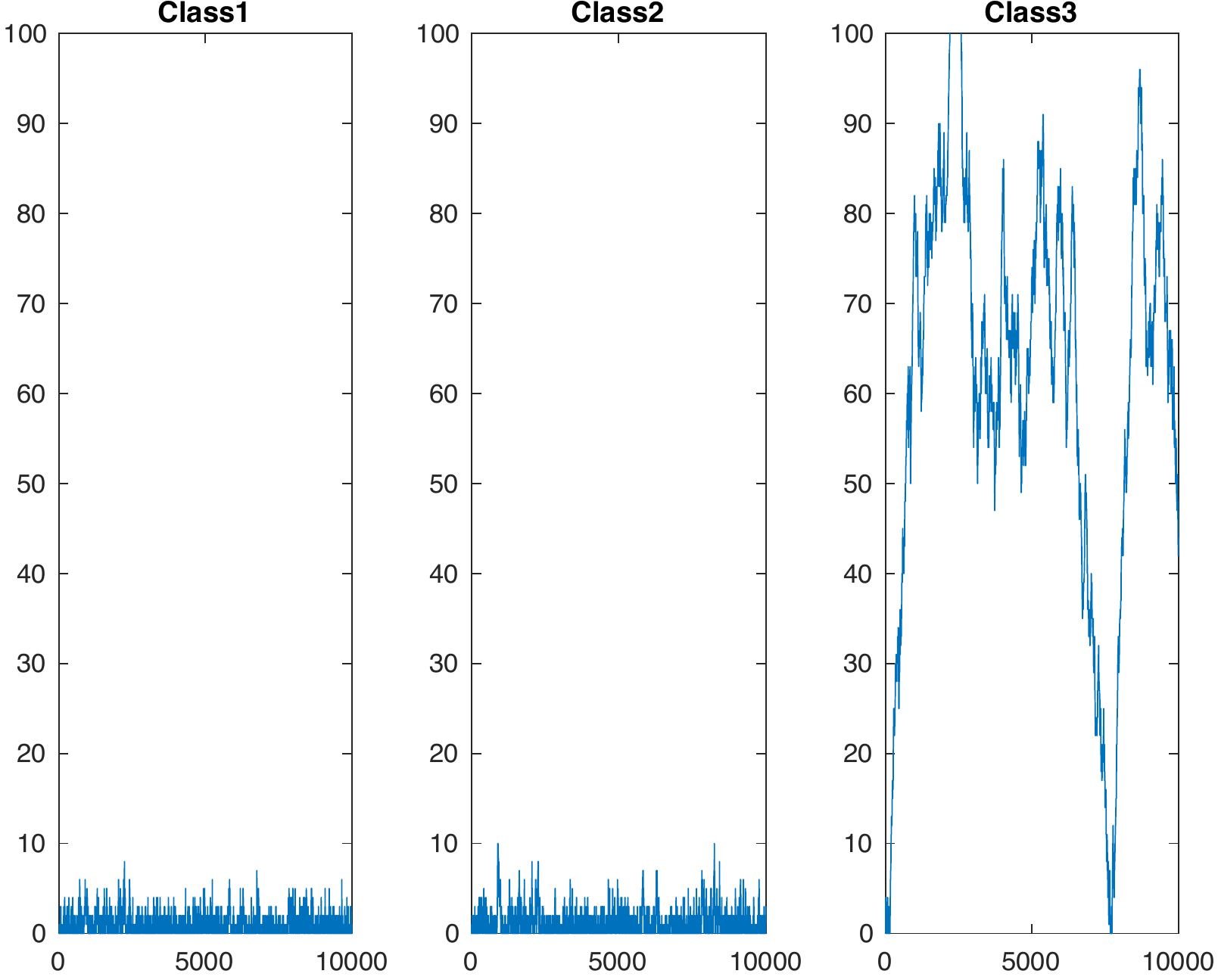}
\caption{Numbers of customers of different classes against time in the system with accumulating but where initial priorities are assigned according to our suggested solution, with $\varepsilon=0.001$}
\label{fig:solution}
\end{figure}

In the case of accumulating priorities one can see (Fig. \eqref{fig:acc}) that the numbers of customers in all classes become large.
If static priorities are used (Fig. \eqref{fig:static}), only the number of customers of class $3$ grows large, whereas the numbers of customers in classes $1$ and $2$ remain reasonably small at all times.
We can also see from Fig. \eqref{fig:solution} below that if our proposed solution is applied, the sample path looks similar to that of the system with static priorities.

\begin{figure}
\centering
\includegraphics[width=0.7\linewidth, height=0.5\linewidth]{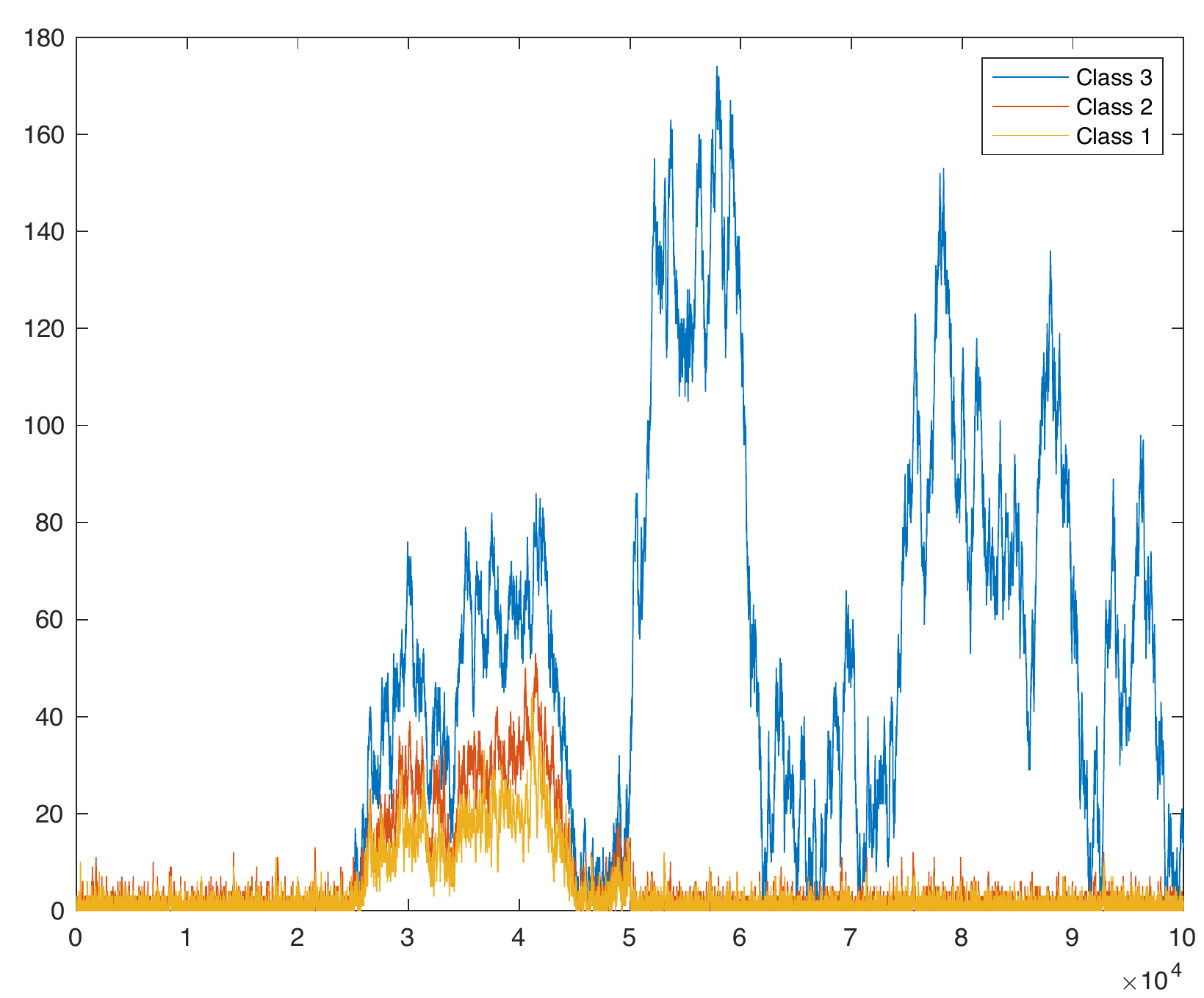}
\caption{Switching from accumulating to static priorities stabilises queues of high-priority customers}
\label{fig:management}
\end{figure}

A further example (see figure \ref{fig:management}) illustrates how by changing the priority regime one can avoid long waiting times for high-priority customers if there is a sudden surge of lower-priority customers. There are, as before, three priority classes, $\lambda_1 = \lambda_2=1/3$ for the entire simulation. In the first quarter of the simulation time $\lambda_3 = 1/3-0.3$, so the total load of the system is $0.7$, the system is in relatively light traffic and the accumulating priorities are used. All queue lengths are small. In the second quarter of the simulation time the arrival rate of the low priority customers suddenly jumps (this could represent for instance seasonal effects) to $\lambda_3=1/3-10^{-3}$, so the total load in the system is $1-10^{-3}$ and the system is in heavy traffic. One can observe all queues becoming large, including that of the highest priority customers. At this point we switch to the static priority regime and apply it for the remainder of the simulation time (arrival rates remain such that the system is in heavy traffic). One can observe that the static priority regime ensures that only the low-priority queue is large, queues of higher-priority customers are small.

\section{Loads above capacity}

In this section we assume that $\Lambda = \lambda_1 + \ldots + \lambda_N < 1$ is fixed. We assume in addition that $\rho = \Lambda + \lambda_{N+1} > 1$ and is also fixed, and further that $\rho - \lambda_i < 1$ for any $i$ (i.e. the system without any class would be stable).  We have two particular cases in mind but do not restrict our attention to these. The first case is similar to the one we had in mind when studying the system in heavy traffic: an increase in the lowest-class arrival rate takes the system load above capacity. Another case may illustrate a catastrophic event, such as for instance a pandemic where a sudden jump in the highest-priority patients may lead to a system operating above capacity.

Since $\rho > 1$, and the system is therefore unstable, in this section we study a fluid version of the model in which we consider separately the queue for each customer class.

We suppose that the level of queue $i$ at time $t > 0$, $L_i(t)$, is given by
$$
L_i(t) = L_i(0) + \lambda_i t - \int_0^t D_i(s) ds,
$$
where $D_i(s)$ denotes instantaneous service rate enjoyed by queue $i$.

\subsection{Static priorities}

In the case of {\bf (SP)} $D_1(s) = \I(L_1(s) > 0) + \lambda_1 \I(L_1(s) = 0)$ for $s > 0$.  As long as $L_1$, the queue for class 1 is strictly positive, all the available service capacity is directed to class 1.  Once $L_1$ has emptied, the new arrivals of class 1 are assigned a dedicated service rate of $\lambda_1$.  Since $\lambda_1 \leq \Lambda < 1$, this guarantees that $L_1(t) = 0$ for all $t \ge T_1$ for some finite $T_1$. 

If $N \geq 2$, then $D_2(t) = 0$ for $t < T_1$.  For values of $t > T_1$, we have $D_2(t) = (1-\lambda_1) \I(L_2(t) > 0) + \lambda_2  \I(L_2(t) = 0$),  that is, a fraction $\lambda_1$ of the available rate is used to keep $L_1$ at zero, and the remaining service capacity is all assigned to queue $2$, while it is positive. Once $L_2$ drops to 0, only a fraction $\lambda_2$ of the available capacity is required to drain the queue length at the same rate at which arrivals occur. Thus, since $\lambda_1 + \lambda_2 \leq \Lambda < 1$, $L_2(t) = 0$ for $t > T_2$ for some finite $T_2$.

Similarly to the above, we can conclude that there exists a finite $T = T_N$ such that $L_1(t) = \ldots = L_N(t) = 0$ for $t \ge T$.  For class $N+1$, on the other hand, $D_{N+1}(t) = 0$ for $t < T$.  When $t \geq T$,  $D_{N+1}(t) = (1-\Lambda) \I(L_{N+1}(t) > 0)$, and thus
$L'_{N+1}(t) = \lambda_{N+1} - (1-\Lambda) = \rho - 1 > 0$, and $L_{N+1}(t) \to \infty$ as $t \to \infty$.

\subsection{Accumulating priorities}

When considering accumulating priorities we define the maximal priority process for each queue $i$, $1 \leq i \leq N+1$, as
$$
P_i(t) = \frac{b_i L_i(t)}{\lambda_i}.
$$
Here we have replaced $W_i(t)$ by $L_i(t)/\lambda_i$, which is the age of the oldest fluid particles in the system -- for the fluid model considered here, these are equivalent.

If $\argmax_j\{P_j(t)\}$ is unique, then 
$$
D_i(t) = \I(i = \argmax_j\{P_j(t)\}),~~~~~~~~~1 \leq i \leq N+1.
$$
On the other hand, if $\argmax_j\{P_j(t)\}$ is not unique, let
$$
\J = \{i: i \in \argmax_j\{P_j(t)\}.
$$
Under the accumulating priority regime, if two or more classes have priority  $\max_j\{P_j(t)\}$ then service capacity should be divided between them in such a way that their priorities remain equal (and maximal). Therefore, if $|\J| > 1$, then $P'_i(t)  = c$, say, for some constant $c \in \mathbb{R}_+$ for all $i \in \J$. Thus
$$
c = P'_i(t) = \frac{b_i}{\lambda_i} (\lambda_i - D_i(t))
$$
and hence
$$
D_i(t) = \lambda_i - \frac{\lambda_i}{b_i} c
$$
for all $i \in \J$. But $\sum_{i \in \J} D_i(t) = 1$ and hence
$$
c = \frac{\sum_{i \in \J}\lambda_i - 1}{\sum_{i \in \J} \lambda_i/b_i}.
$$
Recall that $c = P_i'(t)$ for all $i \in \J$, and recall that we assume $\rho - \lambda_i < 1$ for any $i$. Thus $P_i'(t) < 0$ for any $i \in \J$ as long as $\J$ does not consist of the entire set $\{1,\ldots,N+1\}$ and $P_i'(t) > 0$ as long as $\J = \{1,\ldots,N+1\}$. 

Thus we can now understand the dynamics of the process of priorities: if we start at time 0 with a unique class with the highest fluid priority, its priority is decreasing until it equalises with the priority of another class. From that point onwards, the two priorities stay the same, and both are decreasing at the same rate, until they equalise with the priority of a further class. This continues until all priorities equalise, from which point onwards these priorities grow infinitely. This of course implies that the levels of fluids grow infinitely.

Note also that the above may be summarised for the level of queue $i$ as follows: once all priorities have equalised,
$$
L_i'(t) = \frac{\lambda_i}{b_i} P_i'(t) = \frac{\lambda_i}{b_i} c_i = (\rho-1) \frac{\lambda_i/b_i}{\sum\limits_{j=1}^{N+1} \lambda_j/b_j},
$$
or
$$
\lim_{t \to \infty} \frac{L_i(t)}{t} = (\rho-1) \frac{\lambda_i/b_i}{\sum\limits_{j=1}^{N+1} \lambda_j/b_j},
$$
which shows that the relative queue lengths are exactly as in \eqref{eq:ht_result_queues}.

\subsection{Solution}

As before, a solution to the possibility of queues growing without bound if the accumulating priority regime is applied to all classes is to either employ a static priority regime, or a mixture of accumulating and static priority regimes, but in either case the lowest priority class needs to be operating under the static priority regime.  Both the pure static priority regime, and the mixture, yield identical fluid solutions for classes $1, 2, \ldots, N$ as $t \rightarrow \infty$, with $L_i(t) = 0$, $t > T_S$ for some $T_S > 0$.  On the other hand, $L_{N+1} \rightarrow \infty$ as $t \rightarrow \infty$, under any of the regimes.

\section{Patients' strategic behaviour}
The analysis throughout the previous sections shows the implications of an exogenous growth of demand that loads the system to its (almost) full capacity. As a consequence, depending on the priority regime used, some, or all of the class-specific queues may grow without a bound.

Such an exogenous growth, in the healthcare context, may be a result of a decease outbreak, seasonality, natural disaster, etc. Our findings suggest that the impact of such increase in demand may spread over the whole system. What the analysis so far lacks to capture are secondary effects due to patients behaviour, that is, how patients of different urgency classes adapt to this change. We leave the exact analysis of such an effect to future research. Nevertheless, we provide some insights on this case. 

There is a vast literature on strategic customer behavior in queues. For a comprehensive literature review see \cite{Hassin2003} (until 2003) and \cite{Hassin2016} (from 2004). The most relevant model to the purpose of our analysis is where customers face a ``to queue or not to queue" situation (see, e.g., \cite{Edelson1975}) where customers decide whether or not to join an unobservable queue while taking into account that other customers face the same dilemma. Specifically, assume that each customer incurs a cost of $C>0$ per unit of time they spend in the system and values the service by $R>0$. In the healthcare context, $C$ may be related to the urgency of the patient and $R$ may be seen as the cost and risk associated with no treatment or the price of an alternative (private) treatment that comes with a negligible wait. Here as well, $R$ is related to the urgency of the patient condition where relatively low values of $R$ are associated with elective treatments whereas life threatening condition are associated with extremely high values. 

Given a joining strategy of the other customers, a customer evaluates her expected (stationary) waiting time, henceforth denoted by $W$. Under Nash equilibrium, all the customers use their best-response strategy against the strategy used by the other customers, that is, they decide to join the queue if $CW<R$, to balk if $CW>R$, and join with any probability if $CW=R$. Note that depending on the model under consideration, $C$, $R$, and $W$ may be customer specific.

In the basic model, customers arrive according to a Poisson process at rate $\l$, service times are exponential with rate $\m$, the service regime is first-come first-served (FCFS), and customers are homogeneous with respect to their economic parameters $C$ and $R$. A symmetric (mixed) strategy here is to join the M/M/1 queue with probability $0\le p \le 1$. When all the customers use the strategy $p$, the resulting stationary expected waiting time at the resulting M/M/1 queue is
$$
\frac{1}{\m-p\l}
$$ 
if $\m>p\l$, and $\infty$ otherwise.
If $p$ is such that $C/(\m-p\l)<R$, then an individual customer is better off by joining the queue (with probability 1). Likewise, if $C/(\m-p\l)>R$, a rational customer balks. An individual customer is indifferent between the two options if $R=C/(\m-p\l)$. Therefore, a symmetric Nash equilibrium, in which no customer has the incentive to deviate from the common strategy used by the other customers, is a joining probability $p_e$ such that
$$
p_e= \left\{ \begin{array}{ll}
1 & R> C/(\m-\l),~\m>\l\\
0 & R<C/\m\\
p^* &\mbox{otherwise},
\end{array}
\right.
$$   
where $p^*$ is the unique solution in $p$ of
$$
R= \frac{C}{\m-p\l}.
$$

The analysis above suggests that if $R$ is high enough, that is $R>C/\m$, the equilibrium effective arrival rate as a function of $\l$ equals
 $$
 p^e\l= \left\{ \begin{array}{ll}
\l & \l<\m-C/R\\
\m-C/R & \l \ge \m-C/R, 
\end{array}
\right.
 $$
 and the corresponding equilibrium expected waiting time equals
 \begin{equation}\label{We}
 W^e= \left\{ \begin{array}{ll}
1/(\m-\l) & \l<\m-C/R\\
R/C & \l \ge \m-C/R. 
\end{array}
\right.
\end{equation}
In words, for low arrival rates, the system behaves as an ordinary M/M/1 queue, that is, an increased demand affects expected waiting times. However, past some point, an additional increase of demand will drive the expected waiting past the service value $R$ and some customers will therefore prefer the alternative over joining the queue. The resulting effective arrival rate and expected waiting time are therefore unchanged.

Equation \eqref{We} also suggests that when the potential demand is high enough, the waiting times are in fact determined by the economic parameters $R$ and $C$. In particular, as $R$ in the healthcare context is typically extremely high---especially in life-threatening situations and expensive elective procedures---so are the expected waiting times. Another way to look at this would be that the existence of private alternatives at a reasonable price may act as a safety valve: the private sector absorbs the public system's over-demand when waiting times are going too high.

This basic idea applies also to the multi-class model and priority queues considered in this paper, where the condition and urgency of the patients not only determines their priority class, but also determines their values for $R$ and $C$. The analysis in Section~3 suggests that when the arrival rate grows towards system capacity, the expected waiting time of the lowest priority class (class $N+1$) grows without a bound. Regarding the rest of the priority classes, their waiting times also grow without a bound in the case of AP whereas they remain finite under SP. Taking into account that patients react strategically leads to the following conclusion. If there exist an alternative to class $N+1$ patients at a reasonable price $R$, their effective arrival rate will stop growing at some point, and their, as well as all the other classes' waiting times will remain finite, even when AP are used. Thus, by creating such an alternative (e.g., in the form of reasonably priced private treatment), the effect of high loads are moderated. For more on equilibrium behaviour in multi-class priority queues, see \cite{Haviv2020}.

\section{Discussion}

We have seen that in heavy traffic, the highest priority classes need greater protection than is afforded by the accumulating priority queue with fixed accumulation rates.  This can be achieved either by permitting accumulation rates to grow in inverse proportion to $\epsilon$ in the case $\rho = 1 - \epsilon$, or by applying a static priority regime to the lowest priority class.  In either case the lowest priority class suffers from increasing waiting times, but higher priority classes are protected from this growth.

These results have implications for other scenarios where prioritisation of tasks is a feature.  We discuss below two other important areas of application, but we believe that the potential applications are considerably wider.

Prioritisation of tasks has been introduced in models of human dynamics where, upon completing a task, a person chooses the task from their to-do list with the highest priority to be performed next. A variant of static priorities has been considered in \cite{barabasi2005origin} and a version of accumulating priorities - in \cite{blanchard2007modeling}. Few people would disagree with the observation that at least at some points in our lives we all experience an overload of our to-do lists. This may be modelled as the arrival rate being (perhaps temporarily) close to, or even above, the completion rate, exactly the settings considered in this paper. Our results can therefore be interpreted as follows: when the number of tasks on the to-do list grows, if time-dependent priorities are used, the number of outstanding high-importance tasks will grow. In order to prevent this, either static priorities, or a combination of accumulating and static priorities suggested here, should be used.

Another connection we would like to highlight is to wireless transmission protocols, namely the celebrated MaxWeight introduced in \cite{tassiulas1990stability}. A simple version of it may be described as follows: there are a number of queues, each with its own exogenous stream of arriving jobs, and a single server which, upon completing a job, chooses the next one to perform from the queue with the largest number of outstanding jobs. Other priorities have also been discussed, in particular weighted queue lengths. If one views our model as tasks from the same class forming a queue, then in the case of accumulating priorities the server chooses the next task from the queue with the highest weighted waiting time of the longest-waiting task. Situations considered in this paper are such that the numbers of outstanding tasks in all queues grow to infinity. In this case, the waiting time of the longest-waiting customer is proportional to the number of outstanding tasks. Therefore, in the regimes considered here, the behaviour of the accumulating-priority queue is the same as that of the system governed by an appropriately weighted MaxWeight algorithm.

In this note we focused on average waiting times and queue lengths. It is of course important to study their distributions, which is a subject of our ongoing research. Another research direction we are currently pursuing is a more realistic scenario where customers abandon the system if they waited longer than a certain (perhaps random and perhaps class-dependent) threshold. Strategies minimizing the abandonment rate are of great practical interest.

\bibliographystyle{abbrv}
\bibliography{healthcare}

\end{document}